# ON THE LARGE-$s$ BEHAVIOR OF TOTAL CROSS SECTIONS IN PERTURBATIVE QCD

[Talk at QFTHEP-2000, Tver, 14-20 September 2000]

Fyodor V. Tkachov

INR RAS, Moscow, 117312, Russia
ftkachov@ms2.inr.ac.ru

Reported is a factorization theorem for large-$s$ behavior of total cross sections obtained in a straightforward fashion from ordinary perturbation theory. The theorem extends the standard factorization results for the Bjorken and Drell-Yan processes. The obtaining of useful evolution equations is found to be problematic. A by-product is a direct method for systematic calculations of the parton evolution kernels.

***Disclaimer.***
*The literature on the Regge limit is vast and hard to be familiar with for a newcomer to the field,
so this text is likely to have missed some relevant publications, for which I apologize.
I would appreciate to be notified of a relevant result.*

*Introduction*	1

I am going to report first results of a project in which I attempt to reexamine the problem of large-$s$ behavior of total cross sections in perturbative QCD in a simple-minded but systematic fashion, i.e. starting from ordinary Feynman diagrams and carefully expanding them in the large-$s$ limit.[1] The specific motivations for this project were:

- importance of the subject for the cosmic neutrino physics [2];
- the TOTEM experiment being planned for LHC [3];
- my personal frustration with the non-transparency of the BFKL theory [4] (especially with the concept of reggeized gluons) and many other results of the theoretical school of logarithmic approximations where it is customary to stick a running coupling into a formula by hand.

Since the theory of large-$s$ behavior is closely connected to the theory of deeply inelastic scattering (DIS; see e.g. [5]), an ancillary motivation was

- the problem of calculation of NNLO corrections to the parton evolution (GLAPD [6]) kernels.[2]

Briefly, my conclusion is that an accurate pQCD derivation of the large-$s$ behavior leads to formulae that are a direct extension of the standard results for the Bjorken and Drell-Yan processes. As to the corresponding evolution equation, I have not (yet) found one as straightforward to use in phenomenological applications as the renormalization group equations or the GLAPD equation.

It is far from possible to present complete details in this text due to space-time limitations. However, there seem to be many interesting things one can do based on the result 5.9 (see Sec. 5.13) even without entering into details of its derivation.

*Technical foundation. Asymptotic operation*	1.1

The technical foundation which made this project possible is the theory of asymptotic operation (AO; see the review [8]) — a systematic modern theory of asymptotic expansions of Feynman diagrams.[3]

Asymptotic expansions in parameters is a key piece of any physical formalism. In perturbative quantum field theory, the difficulty of obtaining asymptotic expansions is that formal expansions result in non-integrable singularities in integrands. The theory of AO is based on an efficient algorithmic treatment of the problem of singularities via a powerful mathematical method derived from the theory of distributions [10], [11]. Distributions are singular limits of ordinary functions, and it would have been surprising if they did not emerge in studies of integrands with small parameters. Distributions in this context play an algorithmic role similar to complex or irrational numbers in the theory of algebraic equations. This simple analogy should give one an idea of the inefficiencies implied in the use of the old-fashioned Sudakov-Zimmermann-Lipatov-... methods.

---

[1] This talk is an updated version of [1] where a technical difficulty was overlooked at the final step of derivation (see below Eqs. 5.6, 5.7 and the following discussion). This is corrected in the present version, leading to somewhat less optimistic conclusions about predictability of the large-$s$ behavior within perturbative QCD.

[2] The current state of the art is summarized in [7].

[3] The theory of AO originated in the dark pre-Gorbachev and pre-Internet age behind the Iron Curtain at the fringes of the established theoretical communities [9], which circumstances provided fertile soil for a desinformation of the theoretical community in regard of authorship, originality and theoretical underpinnings of some of the most important theoretical results in perturbative quantum field theory obtained since the early 80s — a desinformation which has been affecting a sizeable fraction of theoretical HEP community and causing a considerable waste (bound to continue into the future) of both research funds and human effort. The brief review of the original papers given below is of course incomplete without a parallel review of "secondary" publications — but ( sorry! no fun this time ☺) space limitations and the focus of this talk do not allow it to branch out in that direction.

The theory of AO extends and generalizes [9] the pattern of reasoning pioneered by Bogolyubov in the discovery [12] of a correct UV renormalization procedure. The variant of AO which treated the class of so-called Euclidean asymptotic regimes was developed in 1982–1988 [9], [13]–[17], [10]. It was a major theoretical breakthrough in several respects (see [8] for more details):

(i) a key concept of perfect factorization (including the requirement of purely power-and-log dependence on the expansion parameter) with far-reaching consequences[4] for both theory and applications of OPE, etc. [13], [15];[5]

(ii) putting Wilson's operator-product expansion on a firm foundation by obtaining it in a form valid for models with zero-mass particles such as QED and QCD [13];

(iii) discovery [13] and simple mnemonic rules [14] for efficient formulae for calculations of Wilson's coefficients in the MS scheme;

(iv) an elegant general algorithmic scheme of derivation of factorization theorems [16], [17] (see below Sec. 3.2);

(v) extension of the results for OPE (including efficient calculational formulae) to arbitrary Euclidean asymptotic regimes, including mass expansions [16], [17];

(vi) a rigorous regularization-independent distribution-theoretic formalism which systematized the method of AO and prepared ground for its extensions (see [10] and refs. therein).

The discovered formulae in combination with the so-called integration-by-parts algorithms [18] formed a theoretical foundation for a flourishing large-scale calculational industry, resulting in an array of NNLO calculations (e.g. [19]; a recent example is [20]).

After the mathematical spadework performed in [10][6] and [22] the extension of AO to non-Euclidean[7] asymptotic regimes required only a simple additional trick of secondary expansion (the so-called homogenization [8], [23][8]). The non-Euclidean AO yielded a complete fully algorithmic solution for the problem of asymptotic expansions of Feynman diagrams summarized in [23]: For any diagram and for any asymptotic regime, the prescriptions of [23] yield the corresponding asymptotic expansion in a maximally simple form which is suitable for obtaining factorization theorems.

The first application of non-Euclidean AO was to the construction of a systematic gauge-invariant perturbation theory for models with unstable fundamental fields [25].[9] This is because the results for amplitudes with only virtual loops can always be rewritten so as to eliminate all traces of distributions[10] and thus pretended to have been derived via ordinary methods (the so-called method of regions), so I had to focus on applications to problems with real phase space[11]. The present project addresses another real phase space problem.

The method of AO can be viewed as a systematic reorganization of the conventional Bogolyubov-Parasyuk-Sudakov-Hepp-Zimmermann-Lipatov-... techniques in such a way as to hide all the complexity of splitting the integration domain into subregions by providing a layer of abstraction based on the notion of singular distribution. AO offers a finite set of rules to generate a power-and-log asymptotic expansion of any Feynman diagram for a given asymptotic regime, with the emphasis not just on splitting the integration domain but also on appropriate modifications of the integrand, depending on the subregion (such modifications may take the form of secondary expansions [8], [23]).

---

[4] Only such expansions attain the ultimate goal of any expansion, which is maximal simplification of calculations; only perfectly factorized results allow a correct phenomenological interpretation and treatment of power corrections; uniqueness of such expansions resulting in an automatic inheritance of gauge identities as well as a drastic simplification of the corresponding analyses (sums and products of such expansions again possess this property).

[5] Earlier versions of OPE — however rigorously proved — were flawed in this respect. See [8] for a discussion.

[6] Construction of a regularization-independent treatment was an important step e.g. in view of the failure of dimensional regularization in some Minkowski-space situations ([21] and Sec. 5 below).

[7] I do not use the term Minkowski-space regimes because Minkowski space allows Euclidean (and quasi-Euclidean) regimes.

[8] The time gap is explained by my involvement with the (then) burning problem of finding the best jet algorithm [24], which prevented me from doing a simple calculation in order to see that the paradox I was puzzled by in 1992 (different results from different scalings) does not exist (because the secondary expansion is done in the sense of distributions [23]).

[9] The work was torpedoed by anonymous referees similarly to [17]. The details are at http://www.inr.ac.ru/~ftkachov/projects/unstable.

[10] This is a technical foundation for the effects mentioned in ftn. 3.

[11] where I'd be more safe from plagiarism if not from anonymous referees.

> AO effectively algebraizes the problem of obtaining asymptotic expansions by offering a higher level of abstraction over — and thus hiding the complexities of — the splittings of integration region which old-fashioned techniques must tackle directly.
>
> AO, therefore, stands in the same relation to old-fashioned techniques as does integral calculus with respect to ancient methods based on explicit summations.

Mathematically, the method of AO is rooted in the theory of distributions (for a discussion see [11]). The distribution-theoretic viewpoint on singularities of perturbation theory was pioneered by Bogolyubov in his studies of UV divergences [12]. This idea (largely ignored by the experts for a quarter of a century) was extended to the problem of asymptotic expansion in the theory of AO [9], with great practical benefits as discussed above.

But the distribution-theoretic power of AO is best seen in the fact that for it, there is no fundamental difference between, say, an ordinary singular function such as $x^{-1}$ and the corresponding singular distribution $\delta(x)$.

> A remarkable consequence of the distribution-theoretic nature of AO is that it works equally well for singular functions such as propagators $(m^2 - p^2 - i0)^{-1}$ and for the related distributions such as $\delta(m^2 - p^2 - i0)$.
>
> This means that the prescriptions of AO [23] are applicable equally well to Feynman diagrams corresponding to amplitudes (virtual loops) and to unitarity diagrams corresponding to matrix elements squared (phase space loops).
>
> The treatment of the two kinds of diagrams is completely uniform with AO, allowing an efficient analysis of real phase-space problems similar to the case of purely virtual loop diagrams.

The old-fashioned methods cannot be applied in such cases without first integrating out $\delta$-functions, resulting in considerable complications due to distortions of the original fundamental product-of-propagators structure of diagrams, which seems to be at least partially responsible for the complexities of the large-*s* theory.

*Setup*  2

Consider the total cross section of the process shown in the picture. We will be having in view standard renormalizable gauge models such as QED and QCD. Masses of the participating fields (electrons or quarks) are collectively denoted as $m$.

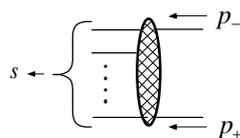

The **Bjorken (DIS) regime** corresponds to the case when $s$ and $Q^2 \equiv -p_-^2$ are both much larger than $p_+^2$ and $m^2$. Componentwise, this is described by $p_\pm - \tilde{p}_\pm = O(m)$, $\tilde{p}_+^2 = 0$, $p_- = O(s^{1/2})$. The quantities pertaining to the asymptotic limit will carry tildes.

The **Regge (large-*s*) regime** corresponds to the case when only $s$ is large: $p_\pm - \tilde{p}_\pm = O(m)$, $\tilde{p}_\pm^2 = 0$.

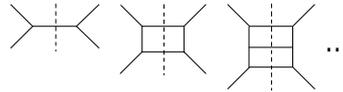 … 2.1

It will be convenient to regard the large parameters as $O(1)$, and expand in the small parameters.

It is also convenient to work with cross sections $\Omega(p_+, p_-, m)$ normalized so as to be dimensionless. In perturbation theory, this is a sum of unitarity diagrams such as shown in Fig. 2.1. The first one is identically zero in the Regge limit (it does however play a role at intermediate steps of the analysis).

My analysis will also involve the quantity $\tilde{\Omega}(\tilde{p}_+, \tilde{p}_-) = \Omega(\tilde{p}_+, \tilde{p}_-, 0)$ — which is usually referred to as the hard parton cross section — with the kinematical parameters formally set to their asymptotic values. Such a quantity is replete with collinear and IR singularities[12] which are manifest in the form of poles in $\varepsilon = \frac{1}{2}(4 - D)$ if dimensional regularization is used. Such singularities correspond to large logarithms in the asymptotic behavior of $\Omega(p_+, p_-, m)$. The method of AO establishes a direct algorithmic connection between such singularities and the large logarithms.

A correct expansion of $\Omega(p_+, p_-, m)$ in the asymptotic regime (see below) yields $\tilde{\Omega}(\tilde{p}_+, \tilde{p}_-)$ with all the singularities/poles appropriately subtracted via special counterterms, so that only logs of $s$ survive.

*One-loop example*  2.2

Consider the simplest box. In the asymptotic limit, its singularities are generated by the two denominators and the two $\delta$-functions, i.e. by the product

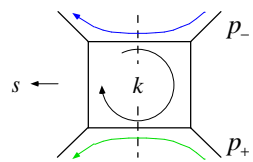

$k^{-2}k^{-2}\delta\left((k+\tilde{p}_+)^2\right)\delta\left((k-\tilde{p}_-)^2\right)$. There is also a numerator (a polynomial of $k$). In the space of $k$, the singularities can be shown as in the following picture.

---

[12] UV singularities play no role and present no complications, and I simply ignore them in the present discussion.

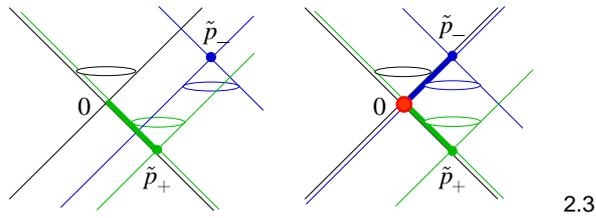

2.3

The left and right figures correspond to the Bjorken and Regge limits, respectively. Each contains three light cones. Their intersections correspond to more complex singularities. The patterns of intersections are related but different: in the Regge case there are more overlaps. Non-integrable singularities are shown with fat lines: green and blue segments correspond to the collinear singularities related to $\tilde{p}_+$ and $\tilde{p}_-$, and described by

$$\begin{cases} k^2 = 0, \\ (k \pm \tilde{p}_\pm)^2 = 0 \end{cases} \Rightarrow k = \mp x\tilde{p}_\pm, \quad x \in (0,1). \qquad 2.4$$

The red dot on the right figure 2.3 corresponds to the soft singularity:

$$\begin{cases} k^2 = 0, \\ (k + \tilde{p}_+)^2 = 0, \Rightarrow k = 0. \\ (k - \tilde{p}_-)^2 = 0 \end{cases} \qquad 2.5$$

Diagrammatically, the factors contributing to the three singularities are as follows:

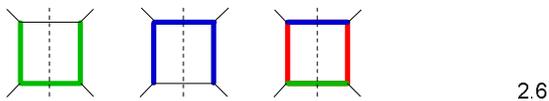

2.6

The green and blue lines correspond to configurations with internal momentum collinear to $\tilde{p}_+$ and $\tilde{p}_-$, respectively. Red lines (the vertical lines in the third figure) are soft (zero momentum). Purely green and blue configurations correspond to GLAPD-type contributions; the third configuration corresponds to a BFKL-type contribution. In the Bjorken limit, only the green configuration contributes, giving rise to the standard result. In the Regge regime, however, the leading logarithmic (L) contribution comes from the third configuration, whereas the collinear contributions are (a) next-to-leading logarithmic (NL) level and (b) proportional to the first diagram in 2.1 which is zero in the Regge limit. So the GLAPD-type contributions actually emerge only at the NNL level.

Neighborhoods of the singularities generate large logarithms in the corresponding limits. The technical problem is to extract those contributions without explicit evaluation of the original diagram. The Sudakov-Lipatov techniques is one way. Asymptotic operation is another.

*Prescriptions of AO*    3

Prescriptions of AO consist of the following steps:

• An accurate formulation of the asymptotic regime in terms of kinematical parameters such as masses and components of external momenta (this we've done already, see Sec. 2).

• Formal (Taylor) expansion of the integrands in the small parameter.

• Enumeration, description, and analysis (power counting) of the singularities of the formal expansion, and connection of the non-integrable ones to the diagrammatic images ("singular subgraphs"). This amounts to construction of the so-called operation $\tilde{R}$ which eliminates non-integrable singularities via additive counterterms that are local in the space of integration momenta. This I will call *renormalization* by analogy with the subtraction of UV divergences. In this case it is usually sufficient to determine a minimal form of the coefficients; if dimensional regularization is applicable, it is sufficient to require that the coefficients contain only pure poles in $\varepsilon$. $\tilde{R}$ generalizes the regularization-independent prescriptions of [10] which in turn are analytically similar to Bogolyubov's *R*-operation in coordinate representation.

• Construction of the so-called bare asymptotic operation $As$ which differs from $\tilde{R}$ only by finite contributions to counterterms. Such finite contributions accumulate the non-analytical dependencies of the final answer (a.k.a. large logarithms of the small parameter) and ensure that the result is a correct asymptotic expansion of the original integral.

• A combinatorial rearrangement (usually called "factorization" although it may actually lead to exponentiation or additive structures) of the contributions from different diagrams for both operations ($\tilde{R}$ and $As$). This can be followed by a derivation of the corresponding evolution equation.

AO establishes a direct correspondence between:

— singular manifolds in the space of integration momenta;

— subsets of singular factors contributing to the corresponding singularities;

— the corresponding diagrammatic images ("subgraphs");

— the countereterms to be added to "bare" expressions. Such counterterms are localized exactly at the singularities of the bare expressions, and are proportional to (derivatives of) $\delta$-functions.

The beauty of the technique is that the writing of $\delta$-functions is mechanical once the singular manifold is parametrized: e.g. for 2.4 the counterterms are[13]

$$\sum_\alpha \int dx \, c_\alpha(x) \, \delta^{(\alpha)}(k \pm x\tilde{p}_\pm), \qquad 3.1$$

where $c$ is a coefficient to be determined.

The least trivial part of the problem reduces to determining the coefficients of counterterms. The systematic prescriptions were summarized in ref. [23]. I only note that an essential element here is the so-called homogenization, or secondary expansion, and that the formalism is kind of fool-proof in the sense that the problem of choosing a correct scaling to do the power counting for complicated singularities does not arise (it is resolved automatically if the prescriptions of AO are followed unwaveringly). Fortunately, sometimes important results can be obtained without writing out explicit expressions for such coefficients.

*Standard Factorization Scenario*  3.2

The general all-logs, all-powers algorithmic scheme of systematic derivation of factorization theorems was pioneered in the theory of AO for OPE and mass expansions [16], [17], and is as follows.

(i) One starts with a quantity to be expanded, say $\Omega(s,m)$, and formulates the asymptotic regime, say, $s \gg m^2$ (here $s$ and $m$ represent the scales of large and small parameters).

(ii) One writes down the so-called "bare" expression $\tilde{\Omega}(s)$ — which is $\Omega(s,m)$ formally (Taylor-)expanded in the asymptotic limit.

(iii) One examines (enumerates and classifies) singularities of the integrand of $\tilde{\Omega}(s)$ in the space of integration momenta.

(iv) One constructs the corresponding minimal renormalization procedure $\tilde{R}$. Represent this symbolically as

$$\tilde{\Omega}_\mu(s) = \tilde{R}\tilde{\Omega}(s) = Z \odot \tilde{\Omega}(s). \qquad 3.3$$

$\tilde{R}$ is applied to integrands, and the operator $Z$ is a "factorized" version of $\tilde{R}$ obtained for the entire PT series. The exact form of $Z$ depends on details of the problem. It may be a multiplicative or additive operation, an integral operator, or any mix of these. The subscript $\mu$ indicates that the usual arbitrariness is involved here, parameterized by $\mu$, as usual.

(v) One constructs the correct asymptotic expansion in the bare form. The point here is that the structure of such an expansion always follows the renormalization 3.3 except that the renormalization factor/kernel now contains, in addition to poles, finite parts with non-trivial (large-logarithmic) dependencies on the small parameters:

$$\tilde{\Omega}_{as}(s,m) = As\,\tilde{\Omega}(s,m) = K(m) \odot \tilde{\Omega}(s). \qquad 3.4$$

The theory of AO guarantees that $\Omega(s,m) \simeq \tilde{\Omega}_{as}(s,m)$ to the desired asymptotic precision, so that the quantity $\tilde{\Omega}_{as}(s,m)$ is what one seeks except for the fact that the two bare quantities on the r.h.s. of 3.4 contain divergences and cannot be directly interpreted in a phenomenological context.

(vi) To this end one constructs an inversion of the renormalization operator $Z$:

$$Z \odot Z^{-1} \equiv 1. \qquad 3.5$$

This important trick was introduced as a tool for deriving renormalization group equations, asymptotic expansions and studies of evolution equations in [28]. The inversion is possible as a formal power series because $Z$ differs from unit operator only by higher-order corrections.

(vii) From 3.5 and 3.4 one obtains

$$\begin{aligned}\tilde{\Omega}_{as}(s,m) &= K(m) \odot Z^{-1} \odot Z \odot \tilde{\Omega}(s) \\ &= K_\mu(m) \odot \tilde{\Omega}_\mu(s).\end{aligned} \qquad 3.6$$

Alternatively, one can construct a renormalization for $K$ and use its inversion. The result is the same up to notations.

(viii) The last step is to obtain evolution equations. To this end one notes that both quantities on the r.h.s. of 3.4 are renormalization group invariant. One simply applies the standard differential renormalization group operator $D = d/d\ln\mu = \partial/\partial\ln\mu + \beta(g)\partial/\partial g$ to 3.3:

$$\begin{aligned}D\tilde{\Omega}_\mu(s) &= DZ \odot \tilde{\Omega}(s) = DZ \odot Z^{-1} \odot Z \odot \tilde{\Omega}(s) \\ &= \gamma \odot \tilde{\Omega}_\mu(s),\end{aligned} \qquad 3.7$$

where the operator $\gamma$ is given by

$$\gamma = DZ \odot Z^{-1}. \qquad 3.8$$

Note that, essentially by construction,

$$D\tilde{\Omega}_{as}(s,m) \equiv 0. \qquad 3.9$$

I called the above Standard Factorization Scenario because: it is the simplest and cleanest way to derive OPE in MS scheme; it worked fine in the discovery of mass expansions and, more generally, expansions for arbitrary Euclidean regimes (see [8] and refs. therein); it also works fine for the Bjorken regime in DIS (as well as the Drell-Yan process away from the threshold).

---

[13] The default lower and upper integration limits are 0 and 1 throughout this text.

*Application to the Bjorken regime*    4

A straightforward application of the above prescriptions gets one the following results. The key point is to write down the explicit form of renormalization 3.3. In this regime (in the leading power approximation but taking into account all logarithms) non-integrable (therefore requiring non-zero counterterms) are only collinear singularities, which are all logarithmic, and one obtains

$$\tilde{\Omega}_\mu(\tilde{p}_+) = \int dx\, x^{-1} Z_{col}(x)\, \tilde{\Omega}(x\tilde{p}_+). \quad 4.1$$

Only the dependence on $\tilde{p}_+$ is shown. The form of this integral operator is predetermined by the structure of the simplest collinear singularity 2.4. The factorization which leads to Eq. 4.1 is depicted as follows:

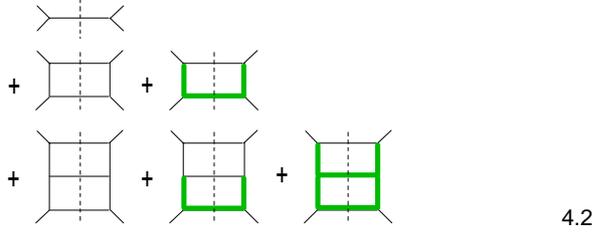

4.2

Comments related to internal loops within subgraphs are given in Sec. 4.8.

From 4.1, one obtains the evolution equation (cf. 3.7):

$$D\tilde{\Omega}_\mu(\tilde{p}_+) = \int_{Q^2/S}^1 dx\, x^{-1} P(x)\, \tilde{\Omega}_\mu(x\tilde{p}_+), \quad 4.3$$

where the lower limit results from the restriction

$$(x\tilde{p}_+ + \tilde{p}_-)^2 = xs - Q^2 > 0. \quad 4.4$$

$P(x)$ is nothing but the GLAPD kernel, and its expression in terms of $Z$ is this (cf. 3.8):

$$P(x) = \int_x^1 dz\, z^{-1} Z_{col}^{-1}(z/x)\, DZ_{col}(z). \quad 4.5$$

The kernel is inverted in the operator sense.

One immediately obtains the AO in bare form (cf. 3.4)

$$\tilde{\Omega}_{as} = \int_{x_{Bjorken}}^1 dx\, x^{-1} K(x,m)\, \tilde{\Omega}(xs), \quad 4.6$$

and its analog in terms of finite quantities (cf. 3.6):

$$\tilde{\Omega}_{as} = \int_{x_{Bjorken}}^1 dx\, x^{-1} K_\mu(x,m)\, \tilde{\Omega}_\mu(xs). \quad 4.7$$

This agrees with the standard DIS factorization. $K_\mu$ is directly seen to be the parton distribution at the scale $\mu$ (trivial summations over parton flavors etc. are omitted).

**Exercise.** Obtain the standard (GLAPD) evolution equation for $K_\mu(x)$ from 4.7, 4.3 and 3.9.

- For clarity's sake: objects like $K_\mu(x,m)$ incorporate, within PT, large logarithms of the small parameter, $\ln(m/\mu)$. Numerical values for such objects cannot be obtained within pQCD and must therefore be treated as phenomenological input determined from experimental data. This was discussed in [15] in connection with the so-called vacuum condensates of local operators. Parton distributions $K_\mu(x,m)$ are another example.

- An interesting related problem is whether the small-$x$ behavior of $K_\mu(x,m)$ can be reliably predicted from PT. Any equation obtained from PT and claimed to predict such a behavior must be based on a strong hypothesis ("behavior at small $x$ obtained from PT persists outside PT even if $K_\mu(x,m)$ is uncalculable from PT at any non-zero $x$" or something essentially equivalent) and must assume a necessarily non-trivial quark-and-gluon model for hadrons.

*On calculability of $P(x)$*    4.8

The explicit expression for $Z_{col}$ has the following structure:

$$Z_{col}(x) = \int dx_1 \ldots dx_l\, \delta(1 - x - \sum x_i)\, \hat{Z}_{col}(x_1, \ldots x_l), \quad 4.9$$

where $\hat{Z}_{col}(x_1, \ldots, x_l)$ is directly connected with the collinear singularities of diagrams contributing to the bare ("hard") cross section. The variables $x_i$ are the fractions of longitudinal momenta of unobserved final state partons (these variables correspond to the cut phase space loops in 4.2). One has

$$\hat{Z}_{col}(x_1, \ldots, x_l) = -\sum_G \mathcal{P}\, \tilde{R}'_{col} G, \quad 4.10$$

where summation runs over all contributing diagrams, $\mathcal{P}$ is the pole part operator, $\tilde{R}_{col}$ is the specific variant of $\tilde{R}$ for collinear singularities, and $\tilde{R}'_{col}$ is $\tilde{R}_{col}$ without the main counterterm. This formula is related to those encountered in the theory of UV renormalization; in the latter case such formulae are written and used in integral sense; in our case Eq. 4.10 directly generalizes the formulae for Euclidean infrared divergences in momentum space (without momentum integrations) given in [10].

In regard of the outstanding problem of NNLO calculation of parton evolution kernels,[14] I can say the following. The calculation of 4.10 is not so much difficult analytically as it is cumbersome due to many parameters (3 in the case of NNLO). In the NLO case, the quantities 4.10 are expressed in terms of simple integrals. In the NNLO case, it is not difficult to obtain simple low-dimensional integrals. Then there remain the integrals in 4.9 and the one in 4.5. In the NLO case these integrals yield results[15] that in my examples are clearly reminiscent of the available answers [26]. In the NNLO case, it is possible to arrive at answers in the form of low-dimensional absolutely convergent integrals (all the poles are extracted at the level of 4.10, so afterwards one only deals with the finite coefficients of such poles) over standard compact regions. Such integrals seem to be easily amenable to numerical integration (I have not studied their analytical calculability). Although rather cumbersome, this scenario does not rely on matching conditions nor on inversion of Mellin transform (as in a comparably cumbersome scenario of [20]). In short, this seems to be a realistic scenario for a *direct* calculation of parton evolution kernels.

---

[14] See [7] for a review of the state of the art.

[15] I am indebted to A. Kotikov for a help with this.

*Application to the Regge regime*     5

The construction of AO for this case has many parallels with the Bjorken case. I first show the analog of 4.2:

                                                   5.1

There is a full complement of terms corresponding to configurations collinear with respect to $\tilde{p}_+$ (purely green; they are the same as in 4.2) as well as a similar set of terms corresponding to $\tilde{p}_-$ (purely blue; they mirror the green ones). Their contribution to $\tilde{R}$ can be immediately written down in the factorized form as follows

$$\tilde{\Omega}_{col}(\tilde{p}_+, \tilde{p}_-) \equiv \tilde{R}_{col} \tilde{\Omega}(\tilde{p}_+, \tilde{p}_-)$$
$$= \int \frac{dx}{x} Z_{col}(x) \int \frac{dy}{y} Z_{col}(y) \tilde{\Omega}(x\tilde{p}_+, y\tilde{p}_-). \quad 5.2$$

There are also qualitatively new terms (the last row; red lines correspond to soft [zero-momentum] singularities). These are a major headache to accurately analyze and compute.[16] Fortunately, the cumulative contribution of such terms after momentum integrations reduce to a single additive contribution:

$$\tilde{\Omega}_\mu(\tilde{p}_+, \tilde{p}_-) = \tilde{\Delta} + \tilde{\Omega}_{col}(\tilde{p}_+, \tilde{p}_-). \quad 5.3$$

The new additive term has no dependence on the external momenta (this is not quite trivial but is the only piece of information from the analysis of singularities that will be used below) and contains only poles in $\varepsilon$.

One immediately writes down the AO in bare form as it must differ from 5.3 only by dependence of renormalization kernels and $\Delta$ on the small parameters (cf. 3.4):

$$\tilde{\Omega}_{as}(\tilde{p}_+, \tilde{p}_-) = \Delta(m)$$
$$+ \int \frac{dx}{x} K(x,m) \int \frac{dy}{y} K(y,m) \tilde{\Omega}(x\tilde{p}_+, y\tilde{p}_-). \quad 5.4$$

Recall that $\tilde{\Omega}$ is the regularized hard parton cross section with all IR and collinear singularities in place.

The form of 5.4 with two convolutions corresponding to the two partons' momenta is meant to bring out the similarity with the standard factorization results for DIS and Drell-Yan processes. From now on I will be using the fact that $\tilde{\Omega}$ are scalar functions of $s = 2\tilde{p}_+ \tilde{p}_-$ only.

The next step is to reexpress $\tilde{\Omega}$ in 5.4 in terms of the finite quantity 5.3. First use inversion of the collinear renormalization similarly to the Bjorken case:

$$\tilde{\Omega}_{as}(s) = \Delta(m) + \int \frac{dz}{z} Q_\mu(z) \tilde{\Omega}_{col}(zs), \quad 5.5$$

where $Q_\mu(z) \equiv \int_z^1 \frac{dx}{x} K_\mu(x,m) K_\mu(z/x,m)$.

$Q_\mu(z)$ is a well-defined function with no stronger than logarithmic singularity near $z = 0$ within PT.

(In the case of the standard Drell-Yan process, there are no soft singularities, $\Delta = 0$ and Eq. 5.5 is equivalent to the standard factorization theorem; see e.g. [5]. Non-zero lower cutoffs for integrals in 5.5 then follow from kinematical restrictions in $\tilde{\Omega}_{col}$.)

*The importance of Eq. 5.5 is that it follows from an accurate analysis of ordinary PT diagrams in the same manner as all the standard factorization theorems. In this respect, Eq. 5.5 is unambiguous.*

To complete the transformation, one is tempted to add and subtract from $\tilde{\Omega}_{col}$ — in order to make it finite (recall 5.3) — the quantity $\tilde{\Delta}$ and rewrite the result as follows:

$$\tilde{\Omega}_{as}(s) \stackrel{?}{=} \Delta_\mu(m) + \int \frac{dz}{z} Q_\mu(z) \tilde{\Omega}_r(xs), \quad 5.6$$

where

---

[16] The subtleties of mixed soft-collinear singularities could be a source of grave mistakes if not handled with care.

$$\Delta_\mu(m) \stackrel{?}{=} \Delta(m) - \tilde{\Delta} \times \left[ \int \frac{dz}{z} Q_\mu(z) \right]. \qquad 5.7$$

If these two formulae made sense then the result would have been extremely attractive:[17] in order to describe the Regge limit within pQCD one would only have to introduce a single new phenomenological constant $\Delta_\mu(m)$; given such a constant and the parton distributions, it would have been sufficient to evolve $\tilde{\Omega}_\mu$ in 5.6.[18]

Alas, there is no reason for the square-bracketed integral to be convergent at $z=0$; in fact, within PT it is logarithmically divergent. So the two expressions 5.6 and 5.7 are ill-defined within PT.

A naive attempt to obtain an evolution equation for $\tilde{\Omega}_\mu$ with respect to $\mu$ similarly to 5.6 also fails: one runs into divergent expressions of the form 5.7 with $Z_{col}$ in place of $K_\mu$, and $Z_{col}(x) \sim O(1)$ near zero in one loop with powers of $\ln x$ in higher orders.

***At this point we've run into a major problem which is the source of all subsequent complications in the study of the large-s regime.***

*The value of the above formalism is that it offers a rather simple explicit formal framework to discuss (if not solve) the problem.*

Let us summarize the relevant properties of 5.5:

- There is no natural kinematic cut as in the Bjorken and Drell-Yan cases, so the integrations over $x$ and $y$ extend all the way down to zero.

- The quantity $\tilde{\Omega}_{col}(s)$ is the hard parton cross section with (only) collinear singularities removed by the collinear renormalization of 5.2, and since the tree level contribution is zero, the singularity at $z=0$ in 5.5 is dimensionally regulated just fine. However, any constant term along with $\tilde{\Omega}_{col}$ causes the integral over $z$ to diverge.[19]

- $K_\mu$ are perfectly well-defined finite parton distributions — exactly the same objects already occurred in the Bjorken regime 4.7. A restriction observed within PT is that the singularity at $x=y=0$ is logarithmic, so that

$$K_\mu(x) \lesssim x^{-\alpha}, \quad \alpha < \tfrac{1}{2}. \qquad 5.8$$

This is the weakest simple hypothesis that could (perhaps ought to, unless its validity is explicitly disproved) be made about parton distributions based on PT.

- The behavior of 5.5 near lower integration limit implicates the non-perturbative parton distributions $K_\mu$. This is a novel feature compared with the conventional situations (Euclidean problems, the Bjorken regime, etc.) where subtractions never require knowledge of non-perturbative dependencies.

Therefore, an important decision one has to make at this point is whether or not one is allowed to make hypotheses about small-*x* behavior of parton distributions.

In fact, it is apparent how to make such hypotheses in a satisfactory fashion given that the $\varepsilon$-dependence of (perturbative) parton distributions is essential for a correct cancellation of poles in 5.5.

If one chooses not to make such hypotheses then there is essentially no choice but to introduce a cut at (low) $z$ in order to split the integration region, and then perform the described identical transformation in the subregion which is not adjacent to zero in order to render $\tilde{\Omega}_{col}$ finite. Obtain:

$$\tilde{\Omega}_{as}(s) = \Delta_{\mu,c}(s) + \int_c^1 \frac{dz}{z} Q_\mu(z) \, \tilde{\Omega}_\mu(zs), \qquad 5.9$$

where

$$\Delta_{\mu,c}(s) = \sum_n \Delta_{\mu,c,n} \ln^n\!\left(s/\mu^2\right). \qquad 5.10$$

Eq. 5.10 is obtained within PT after a complete expansion in $\varepsilon$ of — and canceling the poles (which must cancel by construction) in — the exact expression

$$\Delta(m) + \int_0^c \frac{dz}{z} Q_\mu(z) \, \tilde{\Omega}_{col}(zs) - \tilde{\Delta} \times \int_c^1 \frac{dz}{z} Q_\mu(z). \qquad 5.11$$

The peculiar features of the factorization result 5.9–5.10 are these:

(1) a mixed integral-multiplicative-additive form of the factorization;

(2) an infinite sum in 5.10 which is a function of $s$;

(3) the cutoff $c$ at the lower integration limit.

All this introduces substantial complications in the study of evolution equations.

---

[17] It was reported in [1]. The analysis of [1] was correct up to Eq. 5.5.

[18] After I had reported it in [1], B. Ermolaev notified me of the posting [27] which apparently builds upon an earlier work of the same authors. I am not prepared to comment on [27] due to my unfamiliarity with their formalism which it is a matter of taste to characterize as either more or less cumbersome than mine (it is certainly indirect, inlike mine) — it is just cumbersome enough that I cannot afford to study it in detail immediately or even soon enough. Ref. [27] also claims that a large-*s* (small-*x* in the notations of [27]) pQCD prediction involving only one (if I am not mistaken) phenomenological constant. In view of my findings (see e.g. Eq. 5.10) I am not sure whether it is possible to obtain such a result from pQCD without some hidden hypothesis (or an oversight, as was the case with [1]). I can only express my regret that B. Ermolaev did not offer any comments at the time of [1].

[19] A similar failure of dimensional regularization was found in [21].

*Discussion* 5.12

(i) The derivation of 5.9–5.10 (including the transition from 5.11 to 5.10) rests solely on an accurate study of ordinary PT diagrams. No physical arguments were used (whatever the adjective physical might mean). The non-PT interpretations are made solely after all the expansions, cancellations, etc., leading to 5.9–5.10 are completed.

(ii) All finite quantities depend on $\mu$ as indicated; the expressions are sums of integer powers of $\ln \mu$, as is usual with minimal subtractions. In the case of $\Delta$ and $Q$, the argument of the logarithm actually contains the ratio $\mu/m$ within PT.

(iii) The role of the cutoff $c$ is, in the final respect, similar to that of $\mu$ — and one may even be tempted to choose $c = \mu/s$, perhaps with a simple numerical coefficient. I prefer to keep $\mu$ and $c$ separate to emphasize the fact that whereas $\mu$ corresponds to ordinary PT subtractions, the subtraction associated with $c$ involves a non-perturbative function and is performed in a technically different manner.

(iv) The latter fact implies that along with the evolution in $\mu$, one should study evolution in $c$. I will not discuss the technical problem of how to combine both in a convenient numerical procedure.

(v) To obtain cancellation of poles within PT, one should take into account the $\varepsilon$-dependence of $Q_\mu(z)$. As a result, the coefficients $\Delta_{\mu,c,n}$ receive contributions from $O(\varepsilon^k)$ corrections to perturbative $Q_\mu(z)$, in addition to some weighted integrals of $Q_\mu(z)$ at $\varepsilon = 0$. So the objects $\Delta_{\mu,c,n}$ have to be regarded as new non-perturbative constants in phenomenological interpretations. The latter fact means (if one follows the usual logic) that the expression $\Delta_{\mu,c}(s)$ given by 5.10 is a non-perturbative function to be extracted from experimental data.

Since there is an infinite number of diagrams in the PT series one starts with, there is no a priori principle to prevent appearance of an infinite number of independent objects non-controllable by PT — and this seems to be exactly the situation with the large-*s* regime.

However, the factorization theorem for the Bjorken regime also involves unknown functions (parton distributions) yet it proves possible to obtain meaningful results from it. So it would be premature to draw pessimistic conclusions from the above (although the picture is not rosy either). For instance, it may be important for an efficient phenomenological use of the above factorization to study the (in)dependence of the coefficients $\Delta_{\mu,c,n}$ on the process.

*On the subject of evolution equations* 5.13

There seems to be too many options in regard of derivation of evolution equations, so my purpose here is not to provide a specific algorithmic procedure but to demonstrate the most obvious options.

As discussed above, keeping $\mu$ and $c$ independent means one should study evolution with respect to both parameters. Evolution in $c$ is useful if one chooses $c \propto \mu/s$.

First of all, one could write down two formal evolution equations from the fact that the l.h.s. of 5.9 is independent of $\mu$ and $c$. The differentiation in $c$ yields a simple result:

$$\frac{d}{d\ln c}\Delta_{\mu,c}(s) = Q_\mu(c)\,\tilde{\Omega}_\mu(cs). \qquad 5.14$$

The differentiation in $\mu$ yields

$$D\Delta_{\mu,c}(s) = -\Phi_{\mu,c}(s;K_\mu), \qquad 5.15$$

where $\Phi_{\mu,c}(s;K_\mu) = \int_c^1 \frac{dz}{z} D\left[Q_\mu(z)\tilde{\Omega}_\mu(zs)\right]. \qquad 5.16$

The latter is a functional of the parton distributions which enter via $Q_\mu$. The functional is quadratic in the parton distributions and does not involve their derivatives. It is finite and formally obtainable from PT because the derivative in the integrand is easily calculable: the evolution of parton distributions is known (Sec. 4), whereas $D\tilde{\Omega}_\mu(s)$ is found similarly to 5.9:

$$D\tilde{\Omega}_\mu(s) = 2\int_c^1 \frac{dx}{x} P(x)\,\tilde{\Omega}_\mu(xs) + \delta_{\mu,c}(s), \qquad 5.17$$

where $\delta_{\mu,c}(s) = \sum_{n \geq 1} \delta_{c,n} \ln^n\left(s/\mu^2\right). \qquad 5.18$

The GLAPD evolution kernels $P(x)$ are given by 4.5 and all $\delta_{c,n}$ calculable within PT as power series in the coupling (normalized at $\mu$) with numeric coefficients. Eq. 5.18 is obtained by expansion in $\varepsilon$ of, and canceling poles in, the following expression which is similar to 5.11:

$$\delta_{\mu,c}(s) = D\tilde{\Delta} - 2\tilde{\Delta}\int_c^1 \frac{dx}{x} P(x) + 2\int_0^c \frac{dx}{x} P(x)\,\tilde{\Omega}_{col}(xs). \qquad 5.19$$

Contributions to $\delta_{c,n}$ for larger $n$ start from higher orders of PT but this may not be enough to offset the double-logarithmic nature of the series (log squared per each power of the coupling). So there is a potential problem with convergence here (or a missing exponentiation of the Sudakov type).

Assuming the representation 5.10, we can in principle extract coefficients of $\ln^n(s/\mu^2)$ and obtain evolution equations for the coefficients $\Delta_{\mu,c,n}$:

$$\frac{d}{d\ln c}\Delta_{\mu,c,n} = Q_\mu(c)\,\varphi_n(c)\,, \qquad 5.20$$

$$D\Delta_{\mu,c,n} = (n+1)\Delta_{\mu,c,n+1} - \Phi_{c,n}(K_\mu)\,. \qquad 5.21$$

$\varphi_n(z)$ is a well-defined power series in the coupling which is calculable to any perturbative order. $\Phi_{c,n}(K_\mu)$ is a functional of the parton distribution whose form and coefficients can be found from PT.

One could in principle fix $\mu$ (say, at 10 GeV), take externally provided parton distributions and the constants $\Delta_{\mu,c,n}$, both normalized at the chosen value of $\mu$. Then it would be sufficient to evolve $\tilde\Omega_\mu$ with respect to $s$.

Since $\tilde\Omega_\mu$ is a function of the ratio $s/\mu^2$, the evolution equation in $\mu$ (Eq. 5.17) is in principle (I am not saying "in practice") sufficient.

Let us approach the problem from a different direction. Normalize couplings and parton distributions at $\mu = \sqrt{s}$ so that Eqs. 5.9–5.10 become

$$\tilde\Omega_{as}(s) = \Delta_{\sqrt{s},c,0} + \int_c^1 \frac{dz}{z} Q_{\sqrt{s}}(z)\,\tilde\Omega_{\sqrt{s}}(zs)\,. \qquad 5.22$$

Recall that the perturbative $\tilde\Omega_\mu(s)$ is a sum of $\ln^k(s/\mu^2)$ with coefficients power series in $\alpha_{S,\mu}$, so that $\tilde\Omega_{\sqrt{s}}(zs)$ is a sum of $\ln^k z$ with coefficients power series in $\alpha_{S,\mu=\sqrt{s}}$; the coefficients of the latter power series are just numbers. Then the only quantity that needs to be evolved is $\Delta_{\sqrt{s},c,0}$. If a closed evolution equation for this quantity were available, one would be able to predict the asymptotic cross section 5.22 with just one new phenomenological number. However, Eq. 5.21 couples $\Delta_{\mu,c,n}$ with different $n$.

The form of 5.21 indicates that a diagonalization may be possible. To this end assume the following Ansatz:

$$\Delta_{\mu,c}(s) = \sum_n \Delta'_{\mu,c,n}\ln^n(s/s_0)\,. \qquad 5.23$$

Then Eq. 5.21 becomes

$$D\Delta'_{\mu,c,n} = -\Phi'_{\mu,c,n}(K_\mu)\,. \qquad 5.24$$

The functional on the r.h.s. (which is formally calculable from PT as a coefficient of $\ln^n(s/s_0)$ in 5.16) contains powers of logarithm of the ratio $s_0/\mu^2$. Unfortunately, this diagonalized form cannot be used with 5.22.

- It is not inconceivable that the dynamics of 5.24 (or 5.21) is such that only terms with lowest $n$ in 5.23 (or

5.10) dominate asymptotically, but this requires further analysis.

Remember, however, that all $\Delta_{\mu,c,n}$ are in principle independent constants — independent between themselves and independent from the parton distributions (at least within PT).

Just how useful all the above equations could be, is a question which goes beyond the scope of this text.

## Conclusions 6

A systematic diagram-by-diagram all-logarithms analysis based on the method of asymptotic operation [23], [8] allows one to reproduce the standard factorization theorems and evolution equations for the Bjorken asymptotic regime in DIS and for the Drell-Yan process. The same method was earlier used to find powerful calculational formulae for OPE and mass expansions [13], [16], [17], now in a continuous large-scale use (cf. [20]).

For the large-*s* behavior of total cross sections, the same systematic method yields a factorization theorem (Eq. 5.9) in gauge theories such as QCD and QED. The theorem directly extends the standard results for the Bjorken regime in DIS and for the Drell-Yan processes.

The obtained factorization theorem involves an infinite number of new independent phenomenological coefficients in addition to parton distributions (equivalently, a function which is uncalculable within pQCD). This seems to give a literal interpretation to the characterization[20] of the Regge regime as *infinitely* more complicated than the Bjorken one. I cannot say at this point whether or not this infinite arbitrariness can be reduced in a meaningful fashion (see comments after 5.24).

Whereas I am rather confident that the obtained factorization theorem 5.9 is correct and that no hidden hypothesss was used in its derivation from ordinary PT diagrams, the situation with the corresponding evolution equations is less clear. I have explored some straightforward options for deriving evolution equations, and the conclusions are not encouraging so far: the resulting equations (5.17, 5.21 and 5.20) appear to be rather more involved than the usual renormalization group equations or the GLAPD equations, and it is not clear how to make an efficient and systematic practical use of them.

I am not claiming that the presented formulae constitute a complete formalism to study evolution at large *s* — only the surface has been scratched. There are clearly many options, and determining an optimal way to evolve cross sections to large *s* is a subject for a separate project.

On the up side, the analysis of the Bjorken regime uncovers an algorithmic (although necessarily cumbersome) scenario for direct (i.e. not relying on calculations of moments of structure functions) and systematic (i.e. suitable for next-next-to-leading order) calculations of parton evolution kernels.

---

[20] which I heard in exactly this form from A. White.

Anyhow, the presented formalism offers what seems to be an explicit formal framework to discuss the large-*s* behavior of total cross sections, and some options are likely to have not been explored yet.

Lastly, it is not necessary to enter into details of derivation of the theorems 5.5 and 5.9 in order to play theoretical games with their implications, for which a quite rich field is available (see Sec. 5.13).

*Acknowledgments*

A. Butkevich [2] has been pressing me to help with clarifying the subject of large-*s* behavior of total cross sections for over two years. M. Kienzle-Focacci and W. Kienzle kindly provided information about the TOTEM experiment [3]. Many conversations with L. Lipatov were useful and encouraging, and the inspirational invitation to give a talk at [1] prompted me to sit down to formulae. I thank K. Christos, B. Ermolaev, V. Fadin, M. Grazzini, K. Kato, A. Kotikov, E. Kuraev, A. Vasiljev and A. White for their interest and comments.

This work was supported in part by the Russian Foundation for Basic Research (quantum field theory section) under grant 99-02-18365.